\documentclass[12pt]{article}
\usepackage{graphicx}
\begin{document}

\begin{center}

{\Large \bf Possible Minkowskian Language in Two-level Systems}

\vspace{7mm}

Y. S. Kim\footnote{electronic address: yskim@umd.edu}\\
Department of Physics, University of Maryland,\\
College Park, Maryland 20742

\end{center}

\vspace{5mm}

Pacs: 31.15HZ, 03.30+p, 03.76.Lx
\vspace{5mm}

\begin{abstract}
One hundred years ago, in 1908, Hermann Minkowski completed his
proof that Maxwell's equations are covariant under Lorentz
transformations.  During this process, he introduced a four-dimensional
space called the Minkowskian space.  In 1949, P. A. M. Dirac showed
the Minkowskian space can be handled with the light-cone coordinate
system with squeeze transformations.  While the squeeze is one of
the fundamental mathematical operations in optical sciences, it
could serve useful purposes in two-level systems.  Some possibilities
are considered in this report.  It is shown possible to cross the
light-cone boundary in optical and two-level systems while it is not
possible in Einstein's theory of relativity.
\end{abstract}

\section{Introduction}
Hermann Minkowski was born in Lithuania.  I am very happy to talk about
him in the country where he was born.  He was born in 1864 near the
city of Kaunas.  He then studied at Albertina University in of K\"onigsberg.
Before 1945, K\"onigsberg was the capital city of East Prussia,
located in the Baltic wedge between Lithuania and Poland.  After 1945,
K\"onigsberg became a Russian city of Kaliningrad serving as a Soviet
naval base.

Albertina University, often called the University of K\"onigsberg,
had a very strong academic tradition.  Immanuel Kant was a professor
at this university.  After the total destruction during the second
world war, Soviets started reconstructing the university as the
University of Kaliningrad.  This university is now called
Immanuel Kant State University.

During the 19th century, the University of K\"onigsberg had many
mathematicians interested in physics.  They formulated Maxwell's
equations in the form we use these days.  Minkowski studied
Maxwell's equations there.  Even after he left K\"onigsberg, he
continued his research in Maxwell's equations.  He was a professor
at the University of Zurich while Einstein was a student there.
There he found out that the Maxwell system is covariant under
Lorentz transformations.  Even though Minkowski published his
result in 1908, his interest in Lorentz transformations should
have influenced Einstein who completed his special theory of
relativity in 1905.

Is this the end of the Lorentz completion of the Maxwell system?
Yes, in classical theory of electromagnetism, but No, in the quantum
world.  This problem has a stormy history and was not settled until
1990~\cite{kiwi90jm}.  The question is whether the electromagnetic
field from Maxwell's equations can describe photons as Lorentz-covariant
particles.  I hope to give a review of this subject at another
occasion.

Minkowski also made an strong impact on modern physics through his
Minkowskian space and his geometry.  He introduced space-time system
where
\begin{equation}\label{mscalar}
x^2 + y^2 + z^2 - t^2
\end{equation}
remains invariant under Lorentz transformations.  If the transformation is
made along the $z$ direction, we are dealing with the system where
\begin{equation}
z^2 - t^2 = constant.
\end{equation}
This expression can be written as
\begin{equation}
u v = constant,
\end{equation}
where
\begin{equation}
u = z + t, \qquad v = z - t .
\end{equation}
This means that the Lorentz transformation performs a squeeze
transformation on the $u$ and $v$ variables
\begin{equation}
\pmatrix{e^{\eta/2} & 0 \cr 0 & e^{-\eta/2}}\pmatrix{u \cr v}
= \pmatrix{e^{\eta/2}u \cr  e^{-\eta/2}v} .
\end{equation}
This is a squeeze transformation, where one coordinate is extended
while the other is contracted~\cite{dir49}.

If we make a $45^o$ rotation of this coordinate system, we end up
with $z = (u + v)/2$ and $t = (u - v)/2$, and the transformation
is
\begin{equation}
 \pmatrix{z \cr t} = \pmatrix{\cosh(\eta/2) & \sinh(\eta/2)
               \cr \sinh(\eta/2) & \cosh(\eta/2) } ,
\end{equation}
which is a more familiar expression for the Lorentz transformation.
Indeed, the Lorentz transformation is a squeeze transformation.

While the Lorentz transformation is known to be only for Einstein's
relativity, squeeze transformations are everywhere in optical and
engineering sciences.

In Sec.~\ref{ray}, I review the papers I have published with Sibel
Baskal and Elena Georgieva on ray optics~\cite{gk01,bk02,bk03,gk03}.
In Sec.~\ref{twolev}, I discuss possible squeeze effects in
two-level problems, using the same mathematical formalism developed
for ray optics.

\section{Squeeze Transformations in Ray Optics}\label{ray}
In recent years, I have been working with Sibel Baskal and Elena
Georgieva on ray optics~\cite{gk01,bk02,bk03,gk03}.  More
specifically, we have been interested in beam transfer matrices
of the form
\begin{equation}\label{abcd}
M = \pmatrix{A & B \cr C & D} ,
\end{equation}
with unit determinant or $AD - BC = 1, $ and the matrix elements are
real numbers.  This matrix is often called the $ABCD$ matrix in the
literature.

We have been particularly interested in the $ABCD$ matrix applicable to
periodic systems.  For this purpose, we have shown that this matrix
can be written as a similarity transformation one of the following four
matrices~\cite{bk02,gk03}.
\begin{equation}\label{wigmat}
 \pmatrix{\cos\phi & -\sin\phi \cr \sin\phi & \cos\phi }, \quad
   \pmatrix{\cosh\mu & \sinh\mu \cr \sinh\mu & \cosh\mu} , \quad
   \pmatrix{1 & \alpha \cr 0 & 1}, \quad \pmatrix{1 & 0 \cr \beta & 1} .
\end{equation}
We shall use the notation $W$ collectively for these matrices, and we
write them as $W(\phi), W(\mu), W(\alpha), W(\beta)$ respectively.
Then, according to the property of similarity transformation,
\begin{eqnarray}\label{nphi}
&{}& \left(W(\phi)\right)^N = W(N\phi), \quad
              \left(W(\mu)\right)^N = W(N\mu),  \nonumber \\[1ex]
&{}& \left(W(\alpha)\right)^N = W(N\alpha),  \quad
              \left(W(\beta)\right)^N = W(N\beta).
\end{eqnarray}

Depending on where the cyclic process begins, the similarity
transformation matrix can be as simple as
\begin{equation}
S = \pmatrix{e^{-\eta/2} & 0 \cr 0 & e^{\eta/2}} .
\end{equation}
This is a squeeze matrix which expands one coordinate while contracting
the other.

We are now allowed to write the $ABCD$ matrix of Eq.(\ref{abcd})
as
\begin{equation}
          M = S W S^{-1} .
\end{equation}
Thus
\begin{equation}\label{sws}
M^N = \left(S W S^{-1}\right)^N = S W^N S^{-1} .
\end{equation}
We can then use Eq.(\ref{nphi}) to compute this quantity.

The $ABCD$ matrix can now be written as
\begin{equation}
M = \pmatrix{\cos\phi  & -e^{-\eta}\sin\phi \cr
             e^{\eta}\sin\phi & \cos\phi}
\end{equation}
as the similarity transformation for the first matrix of
Eq.(\ref{wigmat}), and we can write similar expressions for
the remaining matrices.

Using the expression of Eq.(\ref{sws}), we have been able to
deal with the cyclic problems in laser cavities and multilayer
optics.

\section{Possible Squeeze Effects in Two-level Systems}\label{twolev}
Historically, our computation started with the languages the nature
speaks.  We have decimal system because each human being has ten
fingers.  Chinese invented the abacus based on the language these
fingers speak.  Abacuses were used extensively in China, Russia,
and many other countries until the end of the 20th century.

When French army developed long-range guns, the artillery men had
to perform speedy calculations.  They noticed that two bars can
perform additions.  Those with logarithmic scales can perform
multiplications.  The computer based on this principle is called
the slide rule.  Slide rules played essential roles in engineering
applications until hand-held calculators became popular in the
1970s.

John Vincent Atanasoff and John von Neumann noticed vacuum
tubes can speak flip-flop language and started computing machines
based on binary mathematics.  This became the electronic computers
we use every day.

Feynman later noticed that the electron spin can also speak a flip-flop
language, but he noted also that the system has an additional
degree of freedom, leading to the idea of much more powerful computer
than the present electronic system can provide~\cite{fey57}.

We are quite familiar with the quantum mechanics of an electron in
a constant magnetic field.  We also know the electron system can be
described by the spin matrices of the form
\begin{equation}\label{pauli}
S_1 = \frac{1}{2}\pmatrix{0 & 1 \cr 1 & 0}, \quad
S_2 = \frac{1}{2}\pmatrix{0 & -i \cr i & 0}, \quad
S_3 = \frac{1}{2}\pmatrix{1 & 0 \cr 0 & -1}.
\end{equation}
We are quite familiar with these matrices.  They are all Hermitian
and generate rotations in the three-dimensional space.  They
satisfy the set of commutation relations
\begin{equation}
\left[S_i, S_j\right] = i\epsilon_{ijk} S_k .
\end{equation}
This simple mathematical system is the language of nature for quantum
computing~\cite{fey57}.  We note that there are three additional
matrices which will complete the set of commutation relations
which generate the group of two-by-two matrices isomorphic to
the Lorentz group, applicable to the four-dimensional Minkowskian
space.  They are
\begin{equation}
K_1 = \frac{1}{2}\pmatrix{0 & i \cr i & 0}, \quad
K_2 = \frac{1}{2}\pmatrix{0 & 1 \cr -1 & 0}, \quad
K_3 = \frac{1}{2}\pmatrix{i & 0 \cr 0 & -i}.
\end{equation}
While the matrices of Eq.(\ref{pauli}) generate rotations, the above
three matrices generate squeeze transformations.

Let us consider a spin system with a constant magnetic field along the
$y$ direction.  The Hamiltonian is then
\begin{equation}\label{mag}
 h\pmatrix{0 & -i \cr i & 0} .
\end{equation}
The physics of this system is simple enough, but if we add
a dissipative field along the $x$ direction with the form
\begin{equation}\label{dissi}
 g \pmatrix{0 & i \cr i & 0} ,
\end{equation}
the total Hamiltonian becomes
\begin{equation}
H = h\pmatrix{0 & -i \cr i & 0} + g \pmatrix{0 & i \cr i & 0}
= i \pmatrix{0 & -h + g \cr h + g & 0} .
\end{equation}

Then the transition matrix becomes
\begin{equation}\label{tmat}
M = \exp{\left[\pmatrix{0 & -h + g \cr h + g & 0}\right]t} .
\end{equation}
If $g = h$, this expression becomes
\begin{equation}
M = \pmatrix{1 & 0 \cr (2h) t & 1} ,
\end{equation}
and
\begin{equation}
M = \pmatrix{1 & -(2h)t \cr 0 & 1} ,
\end{equation}
if $g = -h$.

Otherwise, we propose to compute this quantity by making a Taylor
expansion.  If $h > g$, we can parametrize the Hamiltonian as
\begin{equation}
i\left(h^2 - g^2\right)^{1/2}
\pmatrix{0 & -\sqrt{(h - g)/(h + g)} \cr \sqrt{(h + g)/(h - g)} & 0} ,
\end{equation}
which can be written as
\begin{equation}
 i\omega \pmatrix{0 & -e^{-\eta} \cr e^{\eta} & 0} ,
\end{equation}
with
\begin{equation}
\omega = \left(h^2 - g^2\right)^{1/2} , \qquad
e^{-\eta} =\left(\frac{h - g}{h + g}\right)^{1/2} .
\end{equation}
Thus, the transition matrix of Eq.(\ref{tmat}) takes the form
\begin{equation}
\exp{\left[\omega\pmatrix{0 & -e^{-\eta} \cr e^{\eta} & 0}t\right]} ,
\end{equation}

Let us note that the matrix
\begin{equation}
\pmatrix{0 & -e^{-\eta} \cr e^{\eta} & 0}
\end{equation}
can be written as a similarity transformation
\begin{equation}
\pmatrix{e^{-\eta/2} & 0 \cr 0 & e^{\eta/2}}\pmatrix{0 & -1 \cr 1 & 0}
\pmatrix{e^{\eta/2} & 0 \cr 0 & e^{-\eta/2}}
\end{equation}
Thus, we can write
\begin{equation}
\left[\pmatrix{0 & -e^{-\eta} \cr e^{\eta} & 0}\right]^N
   = \pmatrix{e^{-\eta/2} & 0 \cr 0 & e^{\eta/2}}
\pmatrix{0 & -1 \cr 1 & 0}^N
\pmatrix{e^{\eta/2} & 0 \cr 0 & e^{-\eta/2}}
\end{equation}
It is therefore possible to compute the transition matrix of Eq.(\ref{tmat})
using Taylor expansion, and the result is
\begin{equation}\label{circle}
\pmatrix{\cos(\omega t)  & -e^{-\eta}\sin(\omega t)  \cr
  e^{\eta}\sin(\omega t)  & \cos(\omega t)} .
\end{equation}

If $g$ is greater than $h$, the off-diagonal elements have the same sign
in Eq.(\ref{tmat}), and we can carry out a similar calculation.  The
result is
\begin{equation}\label{hyper}
\pmatrix{\cosh(\lambda t)  & e^{-\eta}\sinh(\lambda t)  \cr
  e^{\eta}\sinh(\lambda t)  & \cosh(\lambda t)} ,
\end{equation}
with
\begin{equation}
\lambda = \left(g^2 - h^2\right)^{1/2} , \qquad
e^{-\eta} =\left(\frac{g - h}{h + g}\right)^{1/2}  .
\end{equation}

In this way, we obtain a complete analytic set of solutions for this
mixed problem, namely the rotation due to a constant magnetic field
along the $y$ direction, and a squeeze effect along the $x$ direction.

What significance does this result have from the computational point
of view?  Historically, the Minkowskian space was introduced in
connection with special relativity, where the Minkowskian scalar
of Eq.(\ref{mscalar}) remains constant.  Thus, were used to be
firmly committed to the world where the light-cone cannot be
crossed.  This becomes translated into the language that a circle
and cannot be translated into a hyperbola analytically.  This is
an old problem existing since the ancient Greek period.

In this paper, we noted that this barrier can be crossed.  We
can go to the hyperbolic geometry of Eq.(\ref{hyper}) from the
circular geometry given in Eq.(\ref{circle}),by adjusting the
physical parameters $h$ and $g$.  It was noted earlier the same
transition is possible in multilayer optics~\cite{gk03}.

This is precisely the mathematical language optical sciences and
two-level system can speak, Einstein's special language cannot.

\end{document}